\documentclass[aps,prb,reprint,twoside,showpacs,preprintnumbers,amsmath,amssymb,citeautoscript]{revtex4-1}

\pdfoutput=1

\usepackage{graphicx}

\usepackage{lmodern}
\usepackage[T1]{fontenc}
\usepackage[final,babel]{microtype}

\usepackage[utf8]{inputenc}

\usepackage{hyperref}
\usepackage{natbib}

\newcommand{\bra}[1]{\left<#1\right|}
\newcommand{\ket}[1]{\left|#1\right>}
\newcommand{\braket}[3]{\left<#1\left|#2\right|#3\right>}

\newcommand{\half}{\frac{1}{2}}

\newcommand{\DL}{\text{DL}}

\newcommand{\Eq}[1]{Eq.~(\ref{#1})}
\newcommand{\Fig}[1]{Fig.~\ref{#1}}

\newcommand{\av}[1]{\left<{#1}\right>}

\newcommand{\bJ}{{\bf J}}

\newcommand{\Score}{S_\text{core}}
\begin{document}

\title{Modeling and simulations of quantum phase slips in ultrathin superconducting wires}

\author{Andreas Andersson}

\author{Jack Lidmar}
\email{jlidmar@kth.se}

\affiliation{%
Department of Theoretical Physics,
KTH Royal Institute of Technology,
AlbaNova, SE-106 91 Stockholm,
Sweden}

\date{\today}

\begin{abstract}
  We study quantum phase slips (QPS) in ultrathin superconducting wires.
  Starting from an effective one-dimensional microscopic model, which
  includes electromagnetic fluctuations, we map the problem to a
  (1+1)-dimensional gas of interacting instantons.
  We introduce a method to calculate the tunneling amplitude of
  quantum phase slips directly from Monte Carlo simulations.
  This allows us to go beyond the dilute instanton gas approximation
  and study the problem without any limitations of the density of QPS.
  We find that the tunneling amplitude shows a characteristic scaling
  behavior near the superconductor-insulator transition.
  We also calculate the voltage-charge relation of the insulating
  state, which is the dual of the Josephson current-phase relation in
  ordinary superconducting weak links.
  This evolves from a sinusoidal form in the regime of dilute QPS to more
  exotic shapes for higher QPS densities, where interactions are important.
\end{abstract}



\maketitle

\section{Introduction}

In ultrathin, effectively one-dimensional, superconducting wires the
supercurrent may be degraded by phase slips, i.e., sudden unwindings
of the phase of the superconducting order parameter.
Thermal phase slips cause dissipation and in principle destroy
superconductivity at any nonzero temperature.
At zero temperature superconductivity may be disrupted instead by
quantum phase slips (QPS) in which the phase unwinding occurs via
quantum mechanical tunneling events.
The possibility of observing QPS has received much attention
recently~\cite{BezryadinNature2000,LauPRL2001,BollingerPRL2008,Pop2010,Arutyunov2012}.
In wires of finite length, incoherent QPS lead to a small resistivity
even in the superconducting state.
As the thickness of the wire is reduced, the QPS become more
frequent and eventually drive the wire to an insulating state.  In
this regime, the QPS tunneling acts coherently to change the ground
state to a state with the Cooper pairs immobilized.  This state is
characterized by a nontrivial voltage-charge relation, which is the
dual of the Josephson current-phase relation occurring in tunnel
junctions or weak links~\cite{MooijNazarovNature2006}.  Evidence for
coherent QPS was recently observed in wires~\cite{Astafiev2012,Peltonen2013}, and
Josephson junction arrays~\cite{Manucharyan2012}.

Superconducting fluctuations in a wire can be described using an
effective action written in terms of the amplitude $\Delta$ and
phase $\phi$ of the superconducting order
parameter~\cite{GolubevZaikinPRB2001,ArutyunovGolubevZaikinPR2008}.
Quantum phase slips correspond to instantons of this action, which
appear as vortex-like configurations of the phase.  In the limit of
dilute QPS the tunneling amplitude is determined by the action of
single instantons, $t_1 \sim e^{-S_\text{QPS}}$.
Phase slips are, however, nonlocal events, since they require a
rearrangement of the phase over a large portion of the wire.  This
motivates the consideration of interactions between QPS, especially in
the regime near the transition, where they become plentiful.

A phase slip is associated with the tunneling of flux
across the wire, which gives rise to a voltage pulse.  It is
for this reason important to include also the fluctuations of the
electromagnetic fields in the model.

The microscopic action of a superconducting wire which includes these
effects is rather
complicated~\cite{GolubevZaikinPRB2001,ArutyunovGolubevZaikinPR2008}.
In this paper, we model QPS by transforming the action into a model, which is
more manageable and suitable for numerical simulations and
approximations.
We also show how the QPS tunneling amplitude may be extracted from
simulation data, and use this to study its dependence on microscopic
parameters.
Near the superconductor-insulator (SI) transition, it obeys a
characteristic length dependence that can be used to locate the
critical parameters.
Finally, we calculate the characteristic response of the wire, in the
insulating regime of coherent QPS, to an imposed charge displacement.
Similar physics occurs also in discrete Josephson junction
chains~\cite{MatveevPRL2002,Rastelli2013,Fazio2002,Pop2010,Manucharyan2012,Ergul2013}.

The paper is organized as follows: In Sec.~\ref{sec:models}, we
map an effective microscopic action to a dual action
describing a (1+1)D gas of interacting QPS. This reformulation is
crucial and lays the foundation for Sec.~\ref{sec:quantum-phase-slip},
where we introduce an approach to calculate the tunneling amplitude
and the low-energy band relation for QPS. Section~\ref{sec:simulations}
summarizes the grand canonical Monte Carlo method we use to simulate
the system described by our dual action. In Sec.~\ref{sec:results}, we
present and discuss results from the simulations.

\section{Models of superconducting wires}
\label{sec:models}

Our starting point is a microscopic effective action for
thin homogenous superconducting wires in the dirty
limit, derived by Golubev and Zaikin~\cite{GolubevZaikinPRB2001}.
The wires are assumed to be made of conventional $s$-wave
superconductors, well described by BCS theory.  Well below the bulk
superconducting temperature the fermions are gapped and may be
integrated out resulting in a purely bosonic action, which is
subsequently averaged over short range disorder, assuming a mean free
path $l$ (typically $~\sim 5-15$nm) much smaller than the
superconducting coherence length of the clean system
$\xi_0 \sim \hbar v_F /\pi \Delta$ ($\sim 0.1 - 1 \mu\mathrm{m}$).
We will limit the present discussion to relatively homogenous wires
with no long range disorder on scales larger than the coherence length
$\xi \sim \sqrt{l \xi_0}$, although such disorder might lead to
interesting phenomena, e.g., suppression of QPS by Aharanov-Casher
interference effects~\cite{MatveevPRL2002} or Cooper pair
localization~\cite{Feigelman2010}.
The result is a phase-only theory with a constant amplitude
$\Delta_0$, except in the QPS cores (of size
$\sim \xi$) where it will have to momentarily vanish.
In particular, the gap is assumed to remain finite across the SI transition,
where phase coherence is lost.  We here assume the absence of strong
pair breaking interactions, which might otherwise close the fermion
gap before the phase fluctuations destroys
superconductivity~\cite{Sachdev2004,*DelMaestro2009}.

\begin{figure}
  \centering
  \includegraphics[width = 8.8cm]{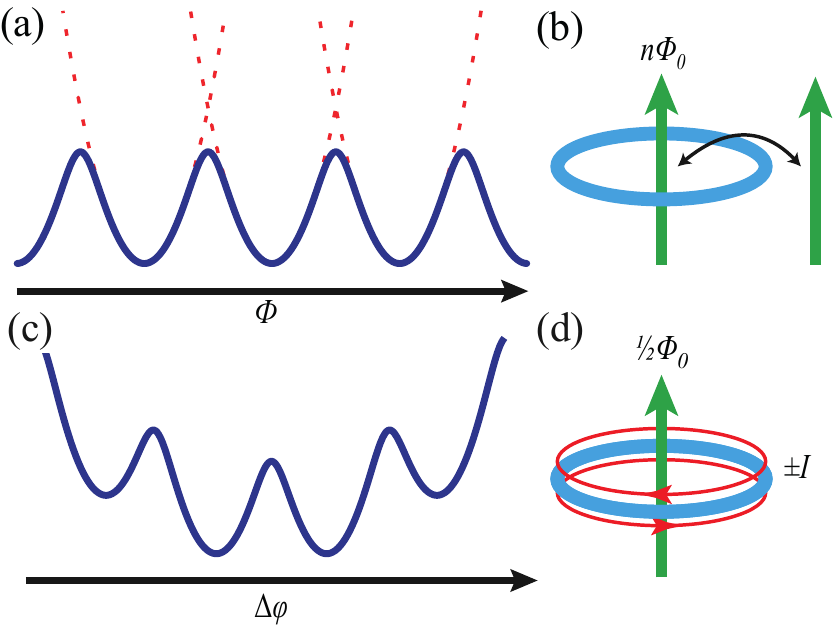}
  \caption{(Color online) (a) Energy vs flux for a closed loop when
    the loop inductance is very large or infinite.  In absence of
    tunneling all states with integer number of flux quanta are
    degenerate, while the tunneling process illustrated in (b) lifts
    this degeneracy.  (c) shows the opposite situation of a small ring
    with small loop inductance, threaded by half a flux quantum.
    There is then just a twofold ground state degeneracy with
    different fluxoid number $n=\oint (d\phi/dx) dx/2\pi$.  The
    tunneling process sketched in (d) connects these two states having
    opposite circulation of the current.  (a), (b) correspond to the
    situation modeled in the present work.}
  \label{fig:flux-tunneling}
\end{figure}

In this section, we map this action, first to a dual charge-current
model, and then to a (1+1)D gas of interacting QPS, paying special
attention to the electromagnetic field.
The resulting model is similar to a 2D Coulomb gas, but with a slightly
more complicated interaction.
We would like to study the QPS occurring in the bulk of the wire and
therefore use periodic boundary conditions to avoid boundary effects.
The geometry of the system is thus a loop, but we will consider it
large enough that we may neglect the effect of the geometric loop
inductance $L_G$, see \Fig{fig:flux-tunneling}(a) and (b).
Under these conditions, a QPS event corresponds to the tunneling of a
flux quantum across the wire, in or out of the loop.
In the limit of very large geometric loop inductance $L_G \to \infty$,
the magnetic energy  $\Phi^2/2L_G$ becomes negligible and the flux
through the loop may fluctuate in time.
As shown below, this fact also relaxes the neutrality constraint usually
present in a 2D Coulomb gas, and allows configurations with a nonzero
net number of instantons.  This point will be important in what follows.

One may consider the geometry chosen more as a theoretical tool to be
able to extract the bulk properties of a wire.  It is straightforward
to also consider smaller loops with non-negligible loop inductance, or
straight wires with boundaries.  A common setup is, e.g., to consider
a small nanoring threaded by half a flux quantum, see
\Fig{fig:flux-tunneling}(c) and (d)~\cite{MatveevPRL2002,Mooij2005,Semenov2013,Rastelli2013}.
In absence of QPS there are then only two degenerate ground states, with
differing fluxoid states, corresponding to clockwise and anticlockwise
persistent currents.  Such geometries have recently provided evidence
for coherent QPS in
experiments~\cite{Astafiev2012,Peltonen2013,Manucharyan2012}.  The
tunneling amplitude in this geometry is not expected to differ
appreciably from the one calculated below, in which the geometric
loop inductance is considered infinite.

In general we use units where $\hbar = 1$, although in some equations
it is reinstated for clarity.

\subsection{Microscopic effective action}
\label{sec:action}
 
The imaginary time action derived in
Refs.~\onlinecite{GolubevZaikinPRB2001,ArutyunovGolubevZaikinPR2008}
for a wire with cross section $s$ is given by
\begin{align} \label{eq:phaseactionSI}
  S &= \int \frac {d\omega}{2\pi} \frac{dk}{2\pi} \bigg\{
  \frac C {2} V^2 +
  \frac {s \chi_D} 2 \left| E \right|^2 + 
  \frac {s \chi_J} {2 (2e)^2} \left|-i \omega \phi - 2e V \right|^2  \notag \\
  &+ \frac {s \chi_L}{8m^2} \left| ik \phi + 2e A \right|^2 + 
  \frac {s \chi_\Delta} 2 \left| \delta \Delta \right|^2 \bigg\}
  \equiv S_\phi + S_\Delta .
\end{align}
Here $\phi$ and $\Delta = \Delta_0 + \delta \Delta$ are the phase and
amplitude of the superconducting order parameter $\psi=\Delta e^{i\phi}$, respectively.
Importantly, the action includes also the electromagnetic field
fluctuations via the vector potential $A$, the voltage $V$, and the
electric field $E = -ikV - i\omega A$.
The Coulomb interaction is parametrized by the capacitance per length $C$.
The geometric inductance of the wire is in general much smaller than
the kinetic one and has been neglected.
This bosonic action is obtained from an expansion of an ordinary BCS action
around the low temperature mean field solution $\Delta_0$, after the
fermions have been integrated out.  The wire
diameter is assumed to be less than the superconducting coherence
length $\xi \approx \sqrt{D/\Delta_0}$, thereby justifying a purely
one-dimensional description.
The kernels $\chi_i$ are given by~\cite{GolubevZaikinPRB2001,ArutyunovGolubevZaikinPR2008}
\begin{align}
  \chi_\Delta &= 2 N_0 					\label{eq:chi_Delta}
  \left(1 + \frac {\omega^2}{12 \Delta_0^2} + \frac {\pi D k^2} {8\Delta_0} \right) ,\\
  \chi_J &= 2 e^2 N_0				\label{eq:chi_J}
  \left(  1 - \frac {\omega^2}{6 \Delta_0^2} - \frac {\pi D k^2} {8\Delta_0} \right) ,\\
  \chi_L &= 2\pi m^2 N_0 D \Delta_0 ,			\label{eq:chi_L} \\
  \chi_D &= \frac {\pi e^2 N_0 D}{4\Delta_0} ,		\label{eq:chi_D}
\end{align}
in the limit of low $\omega$, $D k^2 \ll \Delta_0$, where $\Delta_0$
is the superconducting gap, $N_0$ the density of states at the Fermi
level, $D = v_F l/3$ is the diffusion coefficient, $v_F$ the Fermi
velocity, and $l$ the mean free path.

The action has two parts, $S = S_\Delta + S_\phi$, describing
amplitude and phase fluctuations, respectively.
The phase action allows both for smooth spin-wave fluctuations and
instantons, which are singular vortex-like configurations, where the phase winds by a
multiple of $2\pi$ as the singularity is encircled. 
The latter ones are responsible for the QPS.
At the center of a QPS the amplitude $\Delta$ of the superconducting
order parameter has to go to zero, thereby providing a short distance cutoff for the
otherwise divergent phase contribution.
The large variation of $\Delta$ inside the core thus effectively
couples the phase and amplitude fluctuations, despite the fact that
they appear uncoupled to quadratic order in \Eq{eq:phaseactionSI}.
The action $S_\text{QPS}$ of such instanton configurations
determines to a large extent the QPS tunneling amplitude.
It has a local core part $\Score$, coming from the depletion of the
amplitude $\Delta$ at the centers, and an outer ``hydrodynamic'' part
$S_\phi$ from the phase variation surrounding the cores.

In what follows, we will estimate these contributions separately,
starting with a reformulation of the phase contribution.
Further on, we will study effects of interactions among the QPS.

\subsection{Dual action}

As a first step let us introduce the new fields $\bJ = (\rho, J)$,
where $-2e \rho$ is the electric charge density and $-2e J$ the
electric current, which should be integrated over in the partition
function. By a Hubbard-Stratonovich transformation, we get
\begin{align} \label{eq:link-current}
  S_\phi =& \int \frac {d\omega} {2\pi} \frac {dk} {2\pi} \bigg\{
    \frac C 2 V^2 +
    \frac {s \chi_D} 2 \left| E \right|^2 +
    \frac {(2e)^2}{2 s \chi_J} \left| \rho \right|^2 +
    \frac {4m^2}{2 s \chi_L} \left| J \right|^2 \notag  \\
    &
    + i \rho^* (-i \omega \phi - 2e V)
    + i J^* (i k \phi + 2e A) \bigg\} ,
\end{align}
where $\rho^*(\omega,k)= \rho(-\omega,-k)$, and similarly for $J^*$.
We then split the phase field $\phi = \phi_r + \phi_s$ into a regular
and a singular piece.  The singular part obeys
\begin{align} \label{eq:phidecomp}
  (\partial_x \partial_\tau - \partial_\tau \partial_x) \phi_s = 2\pi v(x,\tau) ,
\end{align}
where 
\begin{equation}\label{eq:vortex-density}
  v(x,\tau) = \sum_i v_i \delta(x-x_i)\delta(\tau-\tau_i),
  \quad
  v_i = \pm 1
\end{equation}
describes density of the phase slips, i.e., vortex-like configurations
in the phase field $\phi(x,\tau)$.
Integrating over the regular
part $\phi_r$ leads to the constraint $\nabla \cdot \bJ = \partial_x J
+ \partial_\tau \rho = 0$, of charge conservation. This can be resolved
by representing the current as
\begin{equation}
\bJ = (\rho,J) = (-\partial_x q, \partial_\tau q)
\end{equation}
or in Fourier space $\bJ = ( -ik q, -i\omega q)$,
where $q(x,\tau)$ is a dimensionless continuous charge field
representing the charge passing through a point $x$ in the wire,
leading to an action
\begin{align} \label{eq:quasi_charge_action} 
  S_\phi =& \int \frac {d\omega}{2\pi} 
  \frac {dk}{2\pi} \bigg\{ \frac C 2 V^2 + 
  \frac {s \chi_D} 2 \left| E \right|^2 + 
  \frac {(2e)^2} {2 s \chi_J} \left|k q \right|^2
  \\ 
  +& \frac {4m^2} {2 s \chi_L} \left| \omega q \right|^2
  + i q(-\omega, -k) ( 2\pi v(\omega,k) - 2e E(\omega,k)) \bigg\}.
  \notag
\end{align}
Since the action is quadratic in both the voltage $V$ and the vector
potential $A$, these can be integrated out exactly. 
We neglect here any spatial variations in the vector potential $A$,
i.e., we treat it as spatially global but time dependent, $A=A(\tau)$.
Fluctuations in $A$ are thus considered instantaneous, an
approximation which is justified because of the high speed of light
and the length of realistic wires, $c \gg L \Delta / \hbar$.
Performing the integrations over $V$ and $A$,
substituting Eqs.~\eqref{eq:chi_J}-\eqref{eq:chi_D}, and expanding $1
/ \chi_J$ to lowest order for small $\omega$ and $k$ obtains
an action which can be expressed as
\begin{align} \label{eq:LCaction}
 S_\phi & 
    = 
    \int
    \frac{d\omega}{2\pi}\frac{dk}{2\pi}    
    \bigg\{
    \frac {(2e)^2} 2 
    \Big( \tilde L(k) \omega^2
      + \frac 1 {\tilde C(k)} k^2 
    \\ &
      + 2\pi\delta(kL) \frac 1 {C \lambda^2}
    \Big)
    \left| q(\omega,k) \right|^2
     + 2\pi i  q(-\omega,-k) v(\omega,k)
    \bigg\} ,  \nonumber
\end{align}
where we introduced the kinetic inductance $\tilde L(k)$ and an
effective capacitance $\tilde C(k)$ (both per unit length) given by
\begin{align}
  \tilde L(k)
  &= \frac 1 {\pi \sigma \Delta_0 s} \left( 1 + \frac {\pi D k^2}{6 \Delta_0 } \right) ,
  \label{eq:L} \\
  \frac 1{\tilde C(k)}
  &= \frac D { \sigma s } \left(
    1 + \frac {\pi D k^2}{8  \Delta_0}
  \right)
  + \frac 1 C \frac 1 { 1 + \lambda^2 k^2 } ,
  \label{eq:C}  \\
  \lambda^2 &=	\label{eq:lambda}
  \frac {\pi \sigma s}{8 \Delta_0 C},
\end{align}
where $\sigma = 2e^2 N_0 D$ is the normal state Drude
conductivity and $\lambda$ is a screening length for the charge carriers.
The term proportional to $\delta(k)$ in \Eq{eq:LCaction} results from
the integration over the spatially constant vector potential $A(\omega)$.

For realistic parameters of experimental wires, the geometric
capacitance per length $C$ is small, $C \ll 2e^2 N_0 s = \sigma s / D$.
Accordingly, we drop the first term in \Eq{eq:C} in this limit. This
also means that $\lambda \gg \xi \simeq \sqrt{D/\Delta_0}$, where $\xi$ is
the coherence length. For low $k \ll \lambda^{-1}$ we get the
velocity of the Mooij-Schön mode~\cite{MooijSchonPRL1985}
\begin{align} \label{eq:c0}
  c_0 = 1/\sqrt{\tilde L C} = \sqrt{\pi\sigma\Delta_0 s/C} .
\end{align}
The terms which are of higher order in $k$ in
Eqs.~\eqref{eq:L}-\eqref{eq:C} become large when $Dk^2/\Delta_0 \sim
1$, i.e., on length scales short compared to $\xi$, which serves as a
short distance cutoff in the model.
Below this cutoff, it is convenient to write the action as
\begin{align}						\label{eq:qaction99}
  S_\phi &= \frac 1 {\beta L} \sum_{k,\omega}
  \frac 1 {2 K c_0} \Big\{
    \left( \omega^2 + \epsilon^2(k) \right) |q(\omega,k)|^2
    \nonumber \\
    & + 2\pi i q(-\omega,-k) v(\omega,k)  \Big\} ,
\end{align}
where we omitted the terms of order $Dk^2/\Delta_0$, and also
introduced infrared cutoffs by considering a wire of finite length with periodic
boundary conditions (i.e., a large loop), at finite temperature
$\beta^{-1}$.
The sum goes over $\omega = 2\pi m / \beta$, $k = 2\pi n /L$, with
$m,n \in \mathbb Z$.
The action has a dispersion
\begin{equation}								\label{eq:epsilon}
  \epsilon(k) = \frac{ c_0 k}{\sqrt{1 + \lambda^2 k^2}}
  + \delta_{k,0} \frac {c_0} {\lambda} ,
\end{equation}
with a coupling constant
\begin{equation}								\label{eq:K}
  K = \frac{\sqrt{\pi \sigma s \Delta_0 C }}{(2e)^2}
  = \frac 1 {(2e)^2} \sqrt{\frac C {\tilde L}}
  = \frac {R_Q} {2\pi Z} ,
\end{equation}
where $\sqrt{\tilde L / {\tilde C}} = Z$ is the impedance of the wire and
$R_Q = h/4e^2$ the resistance quantum.

Although the action in \Eq{eq:qaction99} was derived from a
microscopic theory, a similar action could be written down on
phenomenological grounds, i.e., as the action of a superconducting
transmission line with a linear dispersion modified by the typically
large charge screening length $\lambda$.

In the above equations, we assumed that $\lambda \gg \xi$.  In the
oposite limit $\lambda \ll \xi$, \Eq{eq:qaction99} remains a valid
parametrization of the action, but
with different parameters 
\begin{equation}						\label{eq:tildeparams}
\tilde K = N_0 s \sqrt{\pi D \Delta_0}/2,
\qquad
\tilde c_0 = \sqrt{\pi D \Delta_0},
\end{equation}
and a linear dispersion
$\epsilon(k) = \tilde c_0 k + \delta_{k,0} \Delta_0/\sqrt 8$,
while the intermediate regime $\lambda \sim \xi$ is described by
\Eq{eq:LCaction}.
Note also that for a discrete Josephson junction array one obtains a
similar action, but with 
\begin{equation}						\label{eq:JJparams}
K = \sqrt{E_J/E_{C_0}},
\quad
c_0 = \sqrt{E_J E_{C_0}},
\quad
\lambda = \sqrt{C/C_0},
\end{equation}
where $E_J$ is the Josephson coupling energy, $E_{C_0} = (2e)^2/C_0$,
and $C$, $C_0$ the capacitance of the junctions and to ground,
respectively.

From \Eq{eq:LCaction} or \eqref{eq:qaction99}, one can proceed in
two ways: Sum over the instanton fluctuations $v(x,\tau)$ to
get a sine-Gordon-like action solely in terms of the charge
displacement $q(x,\tau)$, 
\begin{equation}					\label{eq:SG}
  S[q] = S_1[q] - \iint 2y \cos(2\pi q(x,\tau)) \frac{ dx d\tau}{x_0 \tau_0},
\end{equation}
where $S_1$ is the quadratic part of the action (i.e., the first line of
\Eq{eq:qaction99}) and $y$ the QPS ``fugacity'' (see below).
Or one can integrate over the $q$-field to get an effective action for
the quantum phase slips.  We will do the latter here since this gives
a model which can be studied using Monte Carlo simulations relatively
easy.

\subsection{Instanton gas}
\label{sec:gas}
Performing the functional integration over the charge field
$q(x,\tau)$ in \eqref{eq:LCaction} or \eqref{eq:qaction99}
maps the problem to a (1+1)D gas of interacting instantons,
with an action
\begin{align}\label{eq:Sv}
  S_v &= \frac {(2\pi)^2 K c_0} {2 \beta L} \sum_{k,\omega}
 \frac 1 { \omega^2 + \epsilon^2(k) } |v(\omega,k)|^2
\notag \\
&= \half \sum_{\{i,j\}} v_i V(x_i-x_j,\tau_i - \tau_j) v_j,
\end{align}
where $v_i=\pm 1$ denotes the topological charge (vorticity) of
the instantons located at $(x_i,\tau_i)$ and we introduced the pair interaction $V(x,\tau)$.
The summation also needs to be cut off at high $\omega, k$, where the
effective action derived here is no longer applicable. We here use a
Gaussian cutoff, giving an interaction
\begin{align}
  \label{eq:sumVqps}
  V(x,\tau) &= \frac{(2\pi)^2 K c_0  }{\beta L} \sum_{k,\omega} \frac{e^{ikx-i\omega\tau}}{\omega^2 + \epsilon ^2 (k)} 
  e^{-\half (kx_0)^2-\half (\omega \tau_0)^2} .
\end{align}
A natural choice for the ultraviolet cutoffs in space and time are
\begin{equation}
\label{eq:UV-cutoff}
x_0 = \sqrt{\pi D/ 8\Delta_0} \approx \xi,
\qquad
\tau_0 \approx 1/\sqrt 8 \Delta_0,
\end{equation}
so that $\lambda = c_0 \tau_0$.
(This choice also gives
$\tilde c_0=x_0/\tau_0$, $\tilde K=K c_0 / \tilde c_0 = K \lambda / x_0$.)

The problem now has the form of a classical 2D gas of interacting
charges with a partition function
\begin{align} \label{eq:Z}
  Z = \sum_{N=0}^\infty \frac{y^N}{N!} \int
    \prod_{i=1}^N \frac{dx_id\tau_i}{x_0\tau_0} \sum_{v_i=\pm 1} e^{-S_v} . 
\end{align}
The core contribution $\Score$ enters the
fugacity~\cite{GolubevZaikinPRB2001,ArutyunovGolubevZaikinPR2008},
\begin{equation}			\label{eq:fugacity}
  y \approx \Score e^{-\Score}.
\end{equation}
Note that the $k=0$ contribution in \Eq{eq:epsilon} removes the zero mode in
the dispersion and thereby renders the single instanton action $\half
V(0,0)$ finite.
As a result, the partition function includes configurations with no
restriction on the net number of instantons.

\subsection{QPS interaction}

\begin{figure}[]
   \begin{center}
     \includegraphics[width = 8.8cm]{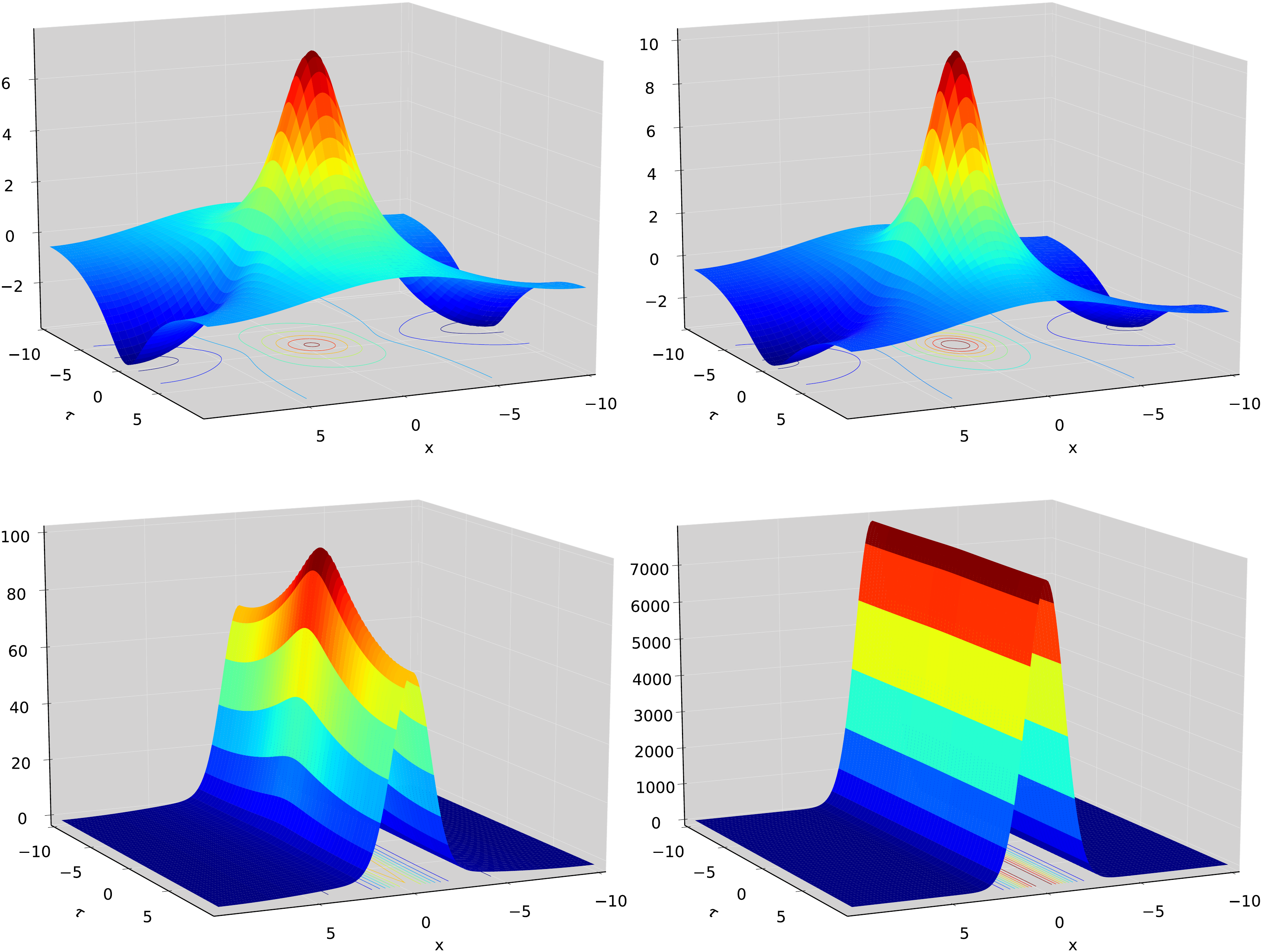}
     \caption{(Color online) The vortex interaction $V(x,\tau)$ for some different values of the parameter $\lambda$. 
       From left to right, top to bottom, the values are $\lambda = $ 0.01, 1.0, 10, and 100. 
       System size is $20 \times 20$, $K=1.0$, and $c_0 = 1$.  Increasing $c_0$ more, squeezes the potential
       in the $\tau$ direction, making it more anisotropic in space-time.
       \label{fig:interaction}
     }
   \end{center}
\end{figure}

The QPS interaction in \Eq{eq:sumVqps} does not allow for an exact
analytic solution, due to the added complication of the charge
screening length $\lambda$. However, on length scales much larger than
$\lambda$ the interaction is purely logarithmic in space-time $V(x,
\tau) \sim \pi K \ln \big[(x^2 + c_0^2 \tau^2)/\lambda^2\big]$, i.e., in
the limit $\lambda \ll x \ll L$, the model is essentially equivalent
to the 2D Coulomb gas. Examples of the general behavior for
finite systems are displayed in \Fig{fig:interaction}, for different
values of $\lambda$.

Of particular interest is the self-interaction
$V_0 = V(0,0)$ in a finite system of length $L$.
In the limit $\beta/\tau_0 \gg 1$, this can be approximated as
\begin{subequations} \label{eq:V00}
\begin{align}
  V_0 &\approx 2 \pi \tilde K \ln (L/x_0), & (L \gg x_0 \gg \lambda) \label{eq:V00a}\\
  V_0 &\approx 2 \pi K  \ln (L/\lambda) + b
  {K\lambda} /{x_0} ,  & ( L \gg \lambda \gg x_0 )
\\
  V_0 &\approx  b K \lambda /{x_0},  & 
  (\lambda \gg x_0, L )
\end{align}
\end{subequations}
 where $b$ is a constant whose precise value depends on the
implementation of the cutoff.  Using \Eq{eq:sumVqps} we estimate $b \approx
0.5 \sqrt{2 \pi^3 } \approx 4$.

\subsection{Core action}
\label{sec:core}

At the center of a QPS, the amplitude $\Delta$ of the superconducting
order parameter has to go to zero, thereby providing a cutoff for the
otherwise divergent phase contribution.  The core part of the
action is
\begin{align}					\label{eq:core-1}
  S_\Delta &= \int \frac{d\omega}{2\pi}\frac{dk}{2\pi}
  \frac{s \chi_\Delta} 2 \left| \delta \Delta \right|^2 \notag \\
  &= \int \frac{d\omega}{2\pi}\frac{dk}{2\pi} N_0 s \left(
    1 + \frac{\omega^2}{12 \Delta_0^2} + \frac {\pi D k^2}{8\Delta_0}
  \right) \left| \delta \Delta \right|^2 .
\end{align}
The typical length and time scales for variations of $\delta\Delta$ are
set by the coherence length $\xi$ and the
inverse gap $1/\Delta_0$.

Following
Refs.~\onlinecite{GolubevZaikinPRB2001,ArutyunovGolubevZaikinPR2008}
we may estimate the core contribution using a Gaussian profile
$\delta\Delta(x,\tau) = \Delta_0 e^{-x^2/2x_0^2 - \tau^2/2\tau_0^2}$,
or in Fourier space $\delta\Delta(\omega,k) = 2\pi \Delta_0 x_0 \tau_0
e^{-k^2x_0^2/2-\omega^2\tau_0^2/2}$.
Carrying out the integration gives
\begin{align} \label{eq:Score}
  \Score = \pi N_0 \Delta_0^2 s x_0 \tau_0
  \left( 1 + \frac 1 {24 \Delta_0^2 \tau_0^2}
    + \frac {\pi D }{16\Delta_0 x_0^2} \right),
\end{align}
and substituting $x_0$, $\tau_0$ by \Eq{eq:UV-cutoff} for consistency,
\begin{equation}					\label{eq:core}
  S_\mathrm{core}
  = a\frac{ K\lambda}{x_0},
\end{equation} 
where $a = 11 \pi/24 \approx 1.44$ is a numerical factor.  The precise
value of $a$ depends on the assumed profile and neglected higher order
terms in the action.
Note that the total cost of inserting one QPS in the system consists of
the core action \Eq{eq:core} and (half) the self-interaction \eqref{eq:V00}.
The latter contains in addition to the logarithmic length dependence a constant
part $\sim b K\lambda/x_0$ with the same parametric dependence on
microscopic parameters as the core part.
Our estimates suggest that they are comparable in magnitude,
$a \approx 1.44$ and $b/2 \approx 2$.
In principle these estimates might be improved on by 
optimizing the total action cost with regard to the
cutoffs~\cite{GolubevZaikinPRB2001,ArutyunovGolubevZaikinPR2008}.
This would, however, only have a minor effect on the numerical
estimates, which are anyhow quite uncertain.
In contrast to previous
work~\cite{GolubevZaikinPRB2001,ArutyunovGolubevZaikinPR2008}, our
estimates of $x_0$ and $\tau_0$ remain parametrically the same also
when $\lambda < x_0$.

In total the local action for one QPS becomes $S_0 = c K \lambda/x_0$,
with $c = a + b/2 \approx 3.44$.
In terms of experimentally accesible parameters,
\begin{equation}					\label{eq:S0}
  S_0 =
  c \sqrt  2 N_0 s \xi \Delta_0 =
  \frac c {4\sqrt 2}\frac{R_Q}{R_N}\frac{L}{\xi},
\end{equation}
where $R_N = L/\sigma s$ is the normal state resistance.

\section{Quantum phase slip amplitude}
\label{sec:quantum-phase-slip}

\subsection{Reduction to zero-dimensional model}
\label{sec:zero-dim}

In this section we will describe a method 
(similar to one employed in a different context in
Refs.~\onlinecite{Vestergren2005,Vestergren2005a})
to calculate the tunneling
amplitude of QPS using the action obtained above.
We will do this by reducing the model to a low energy effectively
zero-dimensional model, treating the whole wire as a lumped element.

As discussed, the instantons or QPS correspond to tunneling events
where the phase slips by $2\pi$.  In order to calculate the tunneling
amplitude of such QPS, it is convenient, although not essential, to
consider a wire with periodic boundary conditions, i.e., a closed
loop.
If we neglect any self-inductance of the loop, the ground-state
energy of the system is periodic in the flux $\Phi$ going
through the loop with period $\Phi_0$, see \Fig{fig:flux-tunneling}.
A quantum phase slip then is accompanied with the tunneling of a flux quantum
$\Phi_0$ in or out of the loop, which will generate a voltage pulse.
Thus we consider the flux $\Phi = L A$ as a dynamical variable.  In the
absence of tunneling, the ground states with $n$ integer flux quanta
are all degenerate.  Tunneling will lift this degeneracy.
At low energies, the whole system can be described by the effective
tight-binding Hamiltonian~\cite{MatveevPRL2002}
\begin{align} \label{eq:tightbinding}
  H &= \sum_n \tilde{E}_0 \ket{n}\bra{n}\notag \\ &- \sum_{n,m} t_m (\ket{n+m}\bra{n} + \ket{n}\bra{n+m}),
\end{align}
where $\tilde{E}_0$ is the ground state energy without tunneling and
$t_m$ are transition amplitudes for the simultaneous tunneling of $m$
flux quanta. Deep in the superconducting phase, where these tunneling
events are rare, the dominating transitions should be for $m = \pm 1$,
but in general also other transition amplitudes may be important~\cite{Weissl2015}.
Due to the translation invariance of the index $n$, the eigenstates of
this Hamiltonian are plane waves:
\begin{align} \label{eq:eigenstates}
  \ket{k} = \frac 1 {\sqrt{N}} \sum_n e^{ikn}\ket{n},
\end{align}
with energies
\begin{align} \label{eq:eigenenegies}
  E_k = \tilde{E}_0 - \sum_{m = 1}^N 2 t_m \cos km, \quad 0 \le k < 2\pi,
\end{align}
where, for normalization purposes, we have limited the number of flux
states to $N$, but eventually this limitation will be  removed.

Our goal here is to relate the tunneling amplitudes $t_m$ to the
properties of the models formulated in the previous section.
A crucial ingredient in those is the inclusion of electromagnetic
fluctuations, which lifts the neutrality constraint in the instanton
gas formulation.
Consider now the partition function $Z_m$ restricted to configurations
where the number of instantons and anti-instantons differ by exactly
$m$  ($m = \sum_i v_i$ is thus the net vorticity).
In the language of quantum mechanics this corresponds to
a matrix element
\begin{align} \label{eq:Zm}
  Z_m = \braket{n+m}{e^{-\beta H}}{n},
\end{align}
which starts from a state $n$ at $\tau = 0$, evolves in imaginary
time, and ends in $n + m$ at $\tau = \beta$.
This allows us to calculate
\begin{align} \label{eq:sumZm_eikm}
  & e^{-\beta E_k} = \braket{k}{e^{-\beta H}}{k} = \frac 1 N \sum_{n,n'}\braket{n'}{e^{-\beta H}}{n}e^{ik(n-n')} \notag  \\
  = &\frac 1 N \sum_{n,m}\braket{n + m}{e^{-\beta H}}{n}e^{ikm} = \sum_m Z_m e^{ikm} .
\end{align}
Denoting $E_k - E_0 = \Delta E_k$, we get
\begin{align} \label{eq:avg_eikm}
  e^{-\beta \Delta E_k}
  = \frac{\sum_m Z_m e^{ikm}}{\sum_m Z_m} = \av{e^{ikm}} .
\end{align}
In a simulation, it is easy to calculate the average $\av{e^{ikm}}$ by first
collecting the histograms $Z_m$ of $m=\sum_i v_i$.
From the low-energy eigenstates $E_k$ calculated using
\Eq{eq:avg_eikm}, we can then obtain the tunneling amplitudes, using
\Eq{eq:eigenenegies}, as
\begin{align} \label{eq:tm}
  t_m = -\int_0^{2\pi} \frac {dk}{2\pi} \Delta E_k e^{-ikm} .
\end{align}
Note that an implicit assumption in this derivation is the
low-temperature limit $\beta \to \infty$.  In practice the $t_m$ and
$E_k$ saturate quickly to constants as $\beta \gtrsim L/c_0$.

\subsection{Voltage-charge relation}
\label{sec:V-k}

The low-energy band relation \Eq{eq:eigenenegies} and its derivatives
are important characteristics of the QPS lumped element.  The $k$ in
this relation corresponds, up to a factor $2e/2\pi$, precisely to an
externally imposed charge displacement $D$.
Indeed, adding a source term 
$i \int D E dx= \int D \dot A dx$ [with
$D = (2e/2\pi) k$] to the action \Eq{eq:phaseactionSI} and going
through the transformations up to \eqref{eq:Sv} leads to an additional
term $i (2\pi/ 2e) D \sum_i v_i = i k m$ in the latter.  $E_k$, 
defined via \Eq{eq:avg_eikm},
therefore gives the energy dependence of the QPS element on an induced
charge.
The first derivative gives the voltage drop
\begin{align} \label{eq:Vk}
  V_k = \frac {2\pi} {2e} \frac {\partial E_k}{\partial k} = \frac {2\pi}
  {2e} \sum_m 2 t_m m \sin km
\end{align}
along the wire as a function of the charge displacement.
The second derivative
\begin{align} \label{eq:Cinvk}
  C_k^{-1} = \left( \frac {2\pi} {2e} \right)^2
\frac {\partial^2  E_k}{\partial k^2} = \left( \frac {2\pi} {2e} \right)^2 \sum_m 2t_m m^2 \cos km
\end{align}
gives the inverse effective capacitance of the wire.
The linear response of the system for $k=0$ is thus capacitative in the
presence of QPS.
Under voltage biased conditions no current will flow below the threshold voltage
$V_{\text{thr}} = \max_k V_k$.

In the regime of coherent QPS, the system is characterized by a
$2e$-periodic voltage-charge displacement relation \eqref{eq:Vk},
which is the dual analog of the Josephson current-phase relation
$I=I_c \sin \gamma$ of the superconducting phase.
When $m=1$ dominates, $V(k)$ reduces to a sinusoidal form, but in
general it may be more complicated.

\subsection{Dilute QPS limit}
\label{sec:DL}

When interactions between the instantons are neglected the method
discussed in Sec.~\ref{sec:zero-dim} reproduces a standard instanton
calculation in the dilute instanton gas approximation.
Let us consider the partition function for phase slips in \Eq{eq:Z} in
this limit and calculate the QPS amplitude.
With $N_+$ instantons and $N_-$ anti-instantons with fugacity ${\tilde y} =
\Score \exp(- \Score - \half V_0)$
 the partition function becomes
\begin{align}
\label{eq:Znonint}
Z = \sum_{N_+,N_-} \frac{{\tilde y}^{N_+}{\tilde y}^{N_-}}{N_+!N_-!} \Omega^{N_+} \Omega^{N_-} = e^{2 {\tilde y} \Omega} .
\end{align}
The variable $\Omega = (L\beta / x_0 \tau_0)$ is the system volume in
space-time. From here, the desired average
$\av{e^{ikm}} = \av{e^{ik(N_+ - N_-)}}$ is easily calculated
\begin{align}
\label{eq:ave^ikm}
\av{e^{ikm}} &= \frac 1 Z \sum_{N_+,N_-} \frac{{\tilde y}^{N_+}{\tilde y}^{N_-}}{N_+!N_-!} \Omega^{N_+} \Omega^{N_-} e^{ik(N_+ - N_-)} \notag \\ &= e^{{\tilde y}\Omega(e^{ik} + e^{-ik}-2)} = e^{-2{\tilde y}\Omega(1-\cos k)} .
\end{align}
This result gives directly
\begin{align}
  \label{eq:Ek}
  \Delta E_k^\DL = \frac{2 L {\tilde y}}{x_0 \tau_0}(1 - \cos k).
\end{align} 
In the dilute instanton limit the QPS amplitude therefore becomes
\begin{align}
\label{eq:t1nonint}
t_1^\DL &= 
 \frac{L {\tilde y}}{x_0 \tau_0} = \frac{L \Score}{x_0 \tau_0} e^{-\Score -
   \half V_0}
 \notag \\
 &\approx 0.72 \Delta_0 \frac L {\xi} \frac{R_Q L}{R_N \xi}
 e^{-\Score - \half V_0} ,
\end{align}
with $\Score$ and $V_0$ given in \Eq{eq:core} and \eqref{eq:V00},
while $t_m=0$ for $m>1$. This implies a purely sinusoidal voltage-charge
relation $V_k = V_c \sin k$, and quantum capacitance $C_k^{-1} \sim \cos k$.
Conversely, a nonsinusoidal relation is a signature of interactions
among the QPS.

\subsection{Superconductor-insulator transition}
\label{sec:SIT}

At low instanton densities the response of the wire is mostly
superconducting with only a small resistance originating from
incoherent QPS.  As the density increases the proliferation of
coherent QPS will drive the wire into an insulating state.
The resulting SI transition has been studied both theoretically and
experimentally in homogenous
nanowires~\cite{BezryadinNature2000,LauPRL2001,BollingerPRL2008,Arutyunov2012,ZaikinPRL1997,ArutyunovGolubevZaikinPR2008,GolubevZaikinPRB2001,RefaelPRB2007,RefaelPRB2009}
and in discrete Josephson
junction chains~\cite{BradleyDoniach84,Fazio2002,Rastelli2013,ChowDelsingHaviland98,Ergul2013}.
An alternative scenario for the SI transition, not directly involving
QPS, has also been proposed for nanowires with strong pair breaking
interactions~\cite{Sachdev2004,*DelMaestro2009}.

The density of QPS is controlled mainly by two parameters appearing in
the instanton action Eqs.~\eqref{eq:V00} and \eqref{eq:core}, the
local contribution $S_0 \sim K \lambda / \xi$ and the
logarithmic term $\sim K \ln L/\lambda$.
In the thermodynamic limit $L \to \infty$, $\beta \to \infty$ the
latter dominates, and may cause the system to undergo a
Berezinskii-Kosterlitz-Thouless (BKT)
transition~\cite{Berezinskii1,*KosterlitzThouless}.
The logarithmic
interaction between the instantons on long length and time scales $x
\gg \lambda$ , $\tau \gg \lambda/c_0 \sim 1/\Delta_0$ is similar to a
2D Coulomb gas.  For high values of $K$ the instantons are all bound
in neutral pairs and the system is in a superconducting state.  As $K$
is decreased below a critical value $K_c = 2/\pi$, the cost for adding
an instanton eventually becomes less than the gain in entropy, and the
proliferation of QPS drives the superconducting wire into an
insulating state.

Note that this quantum phase transition is sharp only in the
thermodynamic limit $L \to \infty$.  For wires of finite length,
several quantities acquire a length dependence characteristic of the
transition.
From \Eq{eq:tightbinding}, the QPS amplitudes $t_m$ have the dimension
of energy.  General scaling arguments at a quantum phase
transition~\cite{Sondhi1997} therefore suggest that $t_1 \sim
L^{-z}$, where $z$ is the dynamic critical exponent.
As the BKT transition is isotropic in space-time, $z=1$.

At very low densities, $t_1$ is given by the approximation
in \Eq{eq:t1nonint}, which can be combined with the expansion of the
self-interaction in the same limit in \Eq{eq:V00} to give
\begin{align} \label{eq:t1scaling}
  t_1^\DL \sim L^{1-\pi K} .
\end{align}
At the critical coupling $K = 2/\pi$ we have $t_1^\DL \sim L^{-1}$,
and so the suggested scaling of $t_1$ holds here.
Going beyond the dilute limit,
interactions among the QPS will renormalize both $K$ and $y$, and the 
actual transition will happen at a $K_c > 2/\pi$ when the fully
renormalized coupling $K_R = 2/\pi$.

Away from the dilute limit the transition can be studied using 
simulations instead. 
The critical value $K_c$ can then be determined from the intersection 
of the curves $Lt_1$ vs $K$, plotted for different $L$ (see below). 

The finite size scaling of $t_1$ near the real transition should, by
an argument similar to Ref.~\onlinecite{Andersson2013}, be subject to a
multiplicative logarithmic correction, so that 
\begin{equation}								\label{eq:t1-logcorr}
  t_1(K_c) \sim L^{-1}/\ln(L/L_0).
\end{equation}
The correction arises because $t_1$ is proportional to the fugacity
$y$ [see \Eq{eq:t1nonint}], which renormalizes logarithmically towards
zero on approaching the fixed point of the transition:
\begin{equation}
  t_1(K_c,y,L) = L^{-z} t_1(K_R,y'(L),1) \sim L^{-1} y'(L).  
\end{equation}

It should be noted that in finite systems various crossovers are
possible, and may in practice mask the real transition.
In particular, the QPS interactions are logarithmic only on length scales
$\gg \lambda$ and thus it requires wires with $L \gg \lambda$.
Although quite uncertain, the screening length can be estimated to be
very large in most experimental wires, perhaps 100--1000 nm, unless
special measures are taken to reduce it.
In this case, the QPS amplitude is dominated by the exponential
dependence on
$S_0 \sim K\lambda/\xi \sim R_Q L/R_N \xi$,
which may still lead to a rather sharp crossover from superconducting
to insulating behavior.

\section{Simulation methods}
\label{sec:simulations}

The instanton gas formulation of Sec.~\ref{sec:gas} is well suited for
simulations.
Since the number of instantons is fluctuating, a grand canonical Monte
Carlo (MC) method is employed.
We use a variant of the scheme developed by Lidmar and
Wallin~\cite{LidmarWallinPRB1997}, which in turn is an extension of
an algorithm by Valleau and Cohen~\cite{ValleauCohenJCP1980}.
In brief, our algorithm consists of five different MC moves: (1)
Creation of a single particle. (2) Destruction of a randomly chosen
single particle. (3) Creation of a neutral pair of particles placed
within a distance $d$ from each other. (4) Destruction of a randomly
chosen neutral pair of particles within a distance $d$ from each other.
(5) Displacement of a randomly chosen particle by a random distance in
the interval $(0,d)$. The maximum displacement and pair distance $d$
is arbitrary and can be tuned to optimize convergence (here we set $d
= 2x_0$).  More details and a derivation of the acceptance ratios for
the creation and destruction moves can be found in
Ref.~\onlinecite{LidmarWallinPRB1997}.

We define an MC sweep to consist of $L \beta/x_0\tau_0$ attempts of any of the four
creation or destruction moves described above, with equal probability
of $1/8$, or the displacement move with probability $1/2$. The system
is typically equilibrated for $10^5$ to $10^6$ MC sweeps, after which
we sample during $10^6$ to $10^7$ sweeps.

To speed up simulations, the QPS interaction in \Eq{eq:sumVqps} is
precalculated on a fine grid and stored in a look-up table. This
enables quick on the fly evaluation by bilinearly interpolating to the
continuous vortex positions.

The tunneling amplitudes $t_m$ are readily calculated from the
histograms of the net topological charge $m$ of the instanton
configurations.
An important technical detail here, is that the exact formula
\Eq{eq:tm} for the QPS amplitude proves to be quite sensitive to
statistical noise, 
especially deep in the superconducting phase when $t_m \ll \Delta_0$.
In those situations, we employ an
alternative method of calculating $t_1$. We extract it from the
effective capacitance, or equivalently from the variance of the
distribution $P_m$ of net instanton number. This variance $\av{m^2}$ is simply
\begin{align}
\label{eq:Pmvar}
\av{m^2} = -\frac {\partial^2}{\partial k^2} \av{e^{ikm}} \Big|_{k=0}
= \beta \sum_{m=1}^\infty 2 t_m m^2.
\end{align}
Assuming $t_1$ dominates the sum, one may approximate it as
$t_1 \approx \av{m^2} / 2\beta$. Although this
estimate of $t_1$ is not exact other than in the dilute limit,
the
distribution often fits nicely to the dilute form $P_m =
e^{-\sigma^2}I_m(\sigma^2)$ ($I_m$ is an $m^{\text{th}}$ order
modified Bessel function and $\sigma = \av{m^2}$), even for quite high QPS densities.
We conclude that the variance should give a reasonable estimate of $t_1$
also in that case.

The tunneling amplitudes $t_m$ are, strictly speaking, ground state
properties of the model and thus require extrapolation to zero
temperature.  All our simulations are carried out in the
low-temperature limit $\beta \gtrsim L/c_0$, where $t_m$ quickly
approaches its zero-temperature limit.
The quantities $E_k$, $V_k$, and $C_k$, which give the response of the
wire to an imposed charge displacement, are meaningful also at nonzero
temperature, although we have not systematically investigated their
temperature dependence.

\section{Results and discussion}
\label{sec:results}
\subsection{QPS amplitude}

A main issue in the field has been the nature of the breakdown of
superconductivity in MoGe nanowires. Early empirical evidence pointed
in the direction of a quantum superconductor-insulator transition
governed by whether the total normal resistance $R_N$ of the wire is
smaller or larger than the resistance quantum
$R_Q$~\cite{BezryadinNature2000}, similar to what happens in a single
Josephson junction shunted by a normal
resistance~\cite{ChakravartyPRL1982,SchmidPRL1983}. However, later
experiments~\cite{LauPRL2001, BollingerPRL2008} suggested that a
better control parameter was the wire cross section area $s$
(proportional to $L/R_N$), at least for wires of length $L \gg 200$
nm.
This is in agreement with the notion that the action for one QPS
$S_0 = \Score + \half V_0$ is the main parameter that determines the
phase boundary.  As discussed above, this is really a crossover, but
since the dependence is exponential it becomes quite sharp.

\begin{figure}[]
  \begin{center}
    \includegraphics[width =8.5cm]{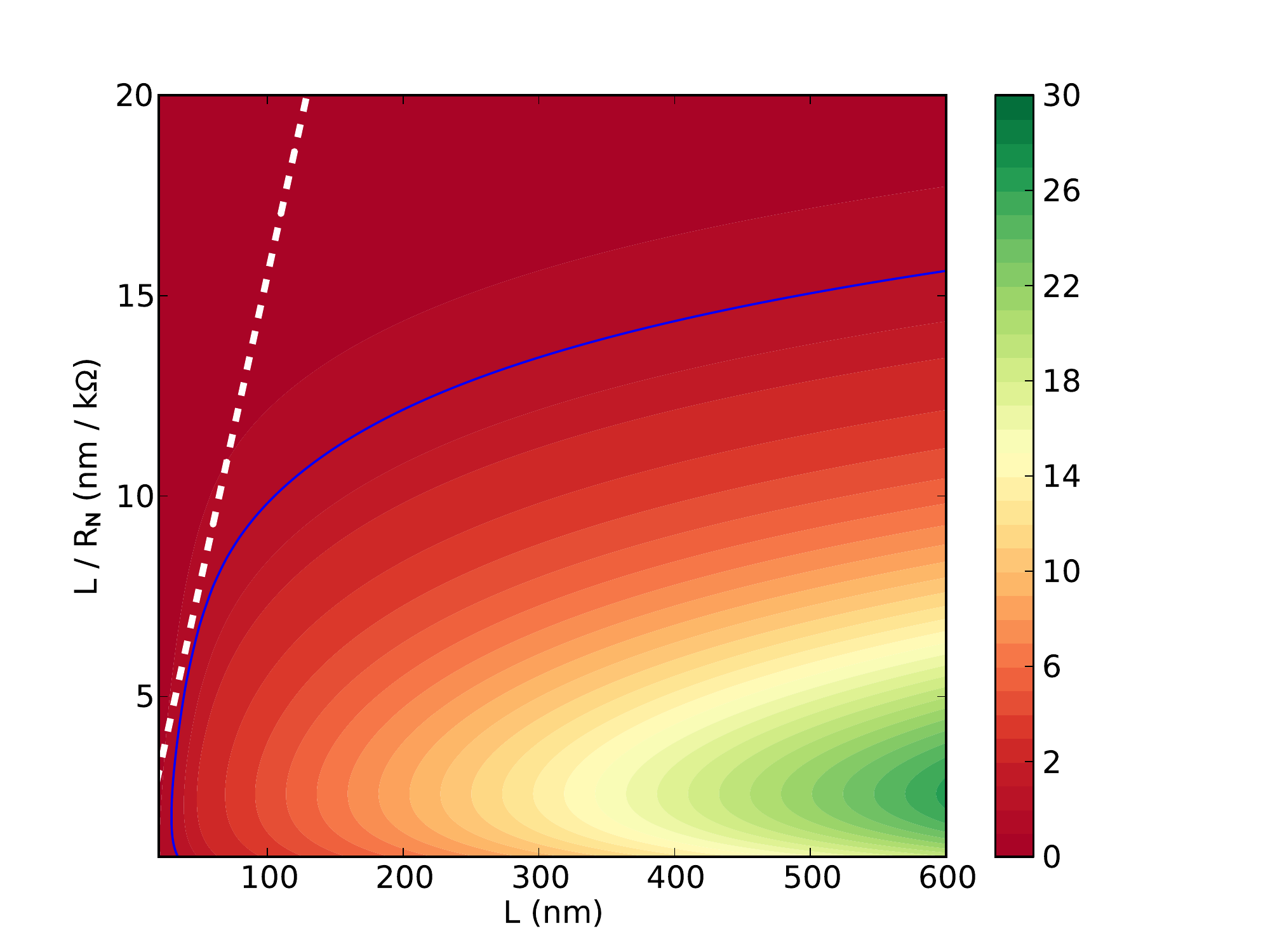}
    \caption{(Color online) The QPS amplitude $t_1^{\DL}$, in the dilute instanton limit,
      in the plane of $L/R_N$ vs $L$.  The QPS amplitude
      is measured in units of $\Delta_0$.  The white dashed line is
      the line $R_N = R_Q$, and the blue line marks the contour $t_1=\Delta_0$.      
      (Assuming $\xi$ = 10 nm, $\Delta_0 = 1.76
      k_B T_c$ (from BCS) with $T_c$ = 5 K, and $C
      \approx $ 5 pF/m.)
      \label{fig:LoverRN_vs_L_t1DL}
    }
  \end{center}
\end{figure}

\begin{figure}[]
  \begin{center}
    \includegraphics[width = 8.5cm]{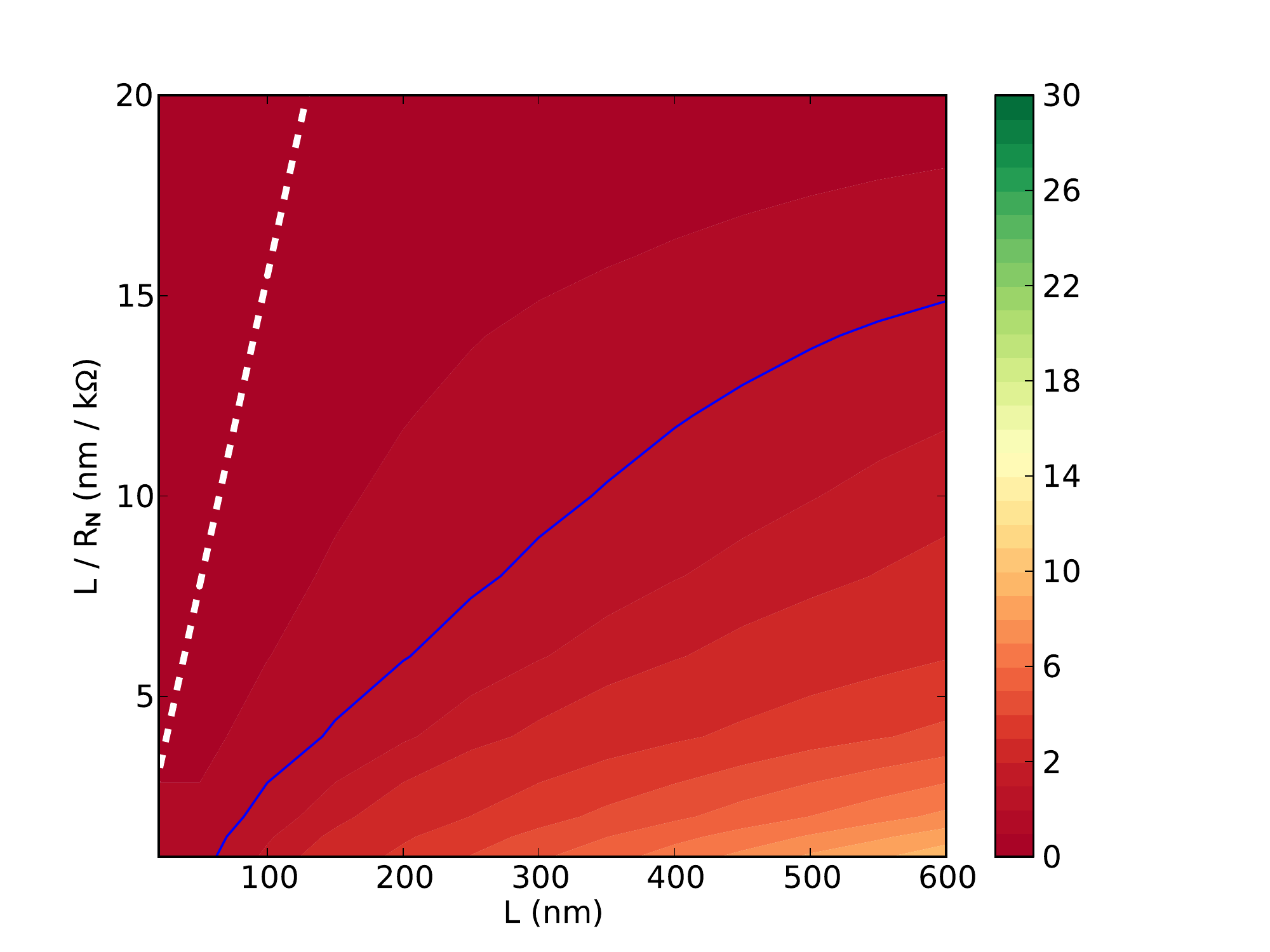}
    \caption{(Color online) The QPS amplitude $t_1$, from simulations, in the plane of $L/R_N$ vs $L$. 
      The QPS amplitude is measured in units of $\Delta_0$.
      The white dashed line is the line $R_N = R_Q$, the blue line
      $t_1 = \Delta_0$.
      \label{fig:LoverRN_vs_L_t1}
    }
  \end{center}
\end{figure}

\begin{figure}[]
  \begin{center}
    \includegraphics[width = 6.5cm]{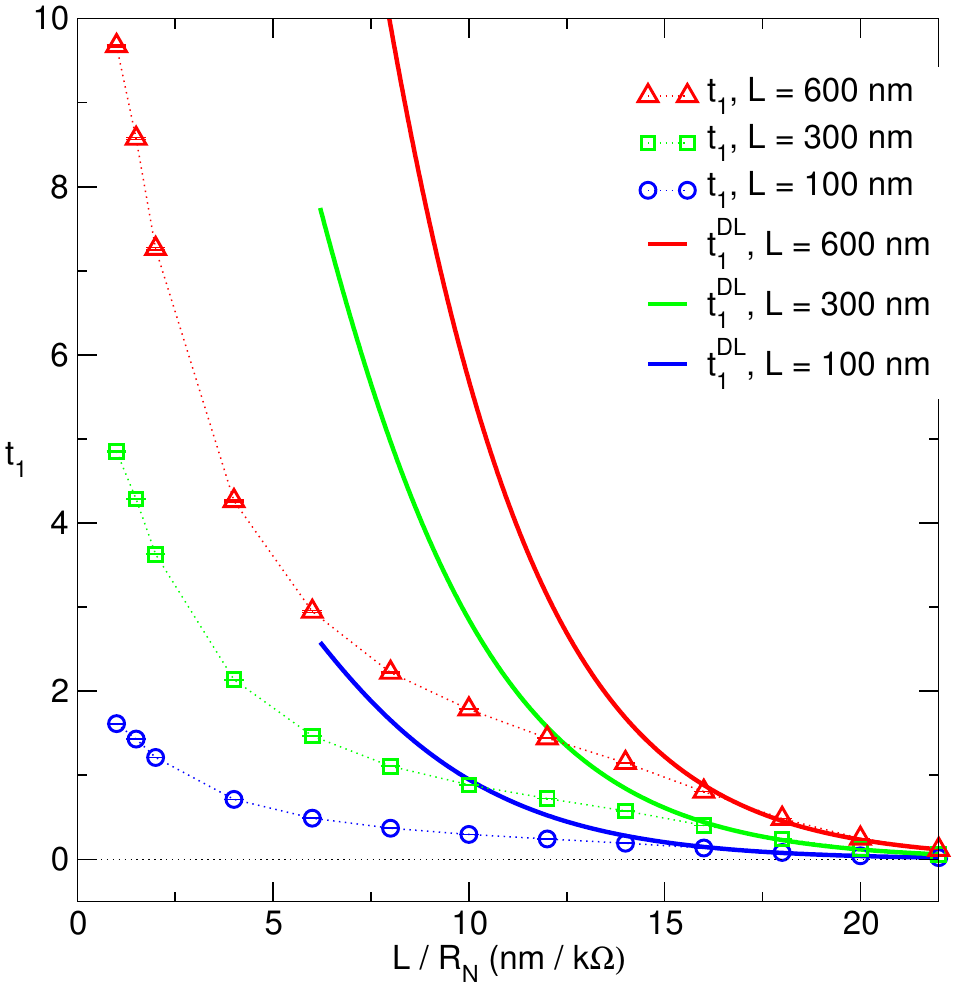}
    \caption{(Color online) The QPS amplitude $t_1$ (from simulations) and $t_1^\DL$ (from the dilute instanton limit) 
      vs $L/R_N$ for fixed $L =$ 100, 300, and 600 nm. The QPS amplitude is measured in units of $\Delta_0$.
      \label{fig:t1_and_t1D_vs_LoverRN}
    }
  \end{center}
\end{figure}

Considering this, we plot in \Fig{fig:LoverRN_vs_L_t1DL} the phase
slip amplitude in the dilute instanton limit, 
\Eq{eq:t1nonint},
in the plane of $L/R_N$ vs $L$. The phase slip amplitude is in this
limit mainly determined by the self-interaction $V_0$, and the core action $\Score$, included in a
chemical potential as 
\begin{equation} \label{eq:mu}
\mu = \ln y = \ln(\Score) -\Score.  
\end{equation}
Including the effects of interactions
between phase slips through MC simulations, the result is that of
\Fig{fig:LoverRN_vs_L_t1}. The phase slip amplitude is now significantly
suppressed for long wires and low $L/R_N$ compared to the dilute case,
but on a qualitative level the two contour plots are quite similar.
[Figure \ref{fig:LoverRN_vs_L_t1DL} displays a nonmonotonic behavior
as function of $L/R_N$, which is absent in the simulation results
shown in \Fig{fig:LoverRN_vs_L_t1}.  This is due to the prefactor in
\Eq{eq:t1nonint} and only occurs in the regime of relatively high QPS
amplitude, where the validity of the dilute limit is highly
questionable.]
In both cases, we take the correlation length $\xi = 10$ nm and the
critical temperature $T_c = 5$ K ($\Delta_0 = 1.76 k_B T_c$), values
typical for MoGe wires~\cite{BezryadinJOP2008}, and also assume a
capacitance of $C \approx $ 5 pF/m.
The difference between $t_1^\DL$ and $t_1$ is best seen in
\Fig{fig:t1_and_t1D_vs_LoverRN}, which shows the same data of
$t_1^\DL$ and $t_1$ vs $L / R_N$, for three different fixed wire
lengths $L =$ 100, 300, and 600 nm. 
The dilute instanton gas approximation consistently overestimates
$t_1$, but as expected, when the QPS amplitude diminishes for larger
$L/R_N$, it becomes better and better.
 
The contour plots in Figs.~\ref{fig:LoverRN_vs_L_t1DL} and
\ref{fig:LoverRN_vs_L_t1} mark the crossover from rare to frequent QPS,
which in practice defines the phase boundary for the
superconductor-insulator transition detected in experiments.  Our
results are in qualitative agreement with the experiments of
Ref.~\onlinecite{BollingerPRL2008}: A steep increase close to
$R = R_Q$ (the dashed white line in Figs.~\ref{fig:LoverRN_vs_L_t1DL}
and \ref{fig:LoverRN_vs_L_t1}) for short wires, which then flattens
out to a more or less horizontal line $L/R_N =$ const. for longer wires.

\subsection{Phase slip interaction effects}
\begin{figure}[]
   \begin{center}
     \includegraphics[width = 7cm]{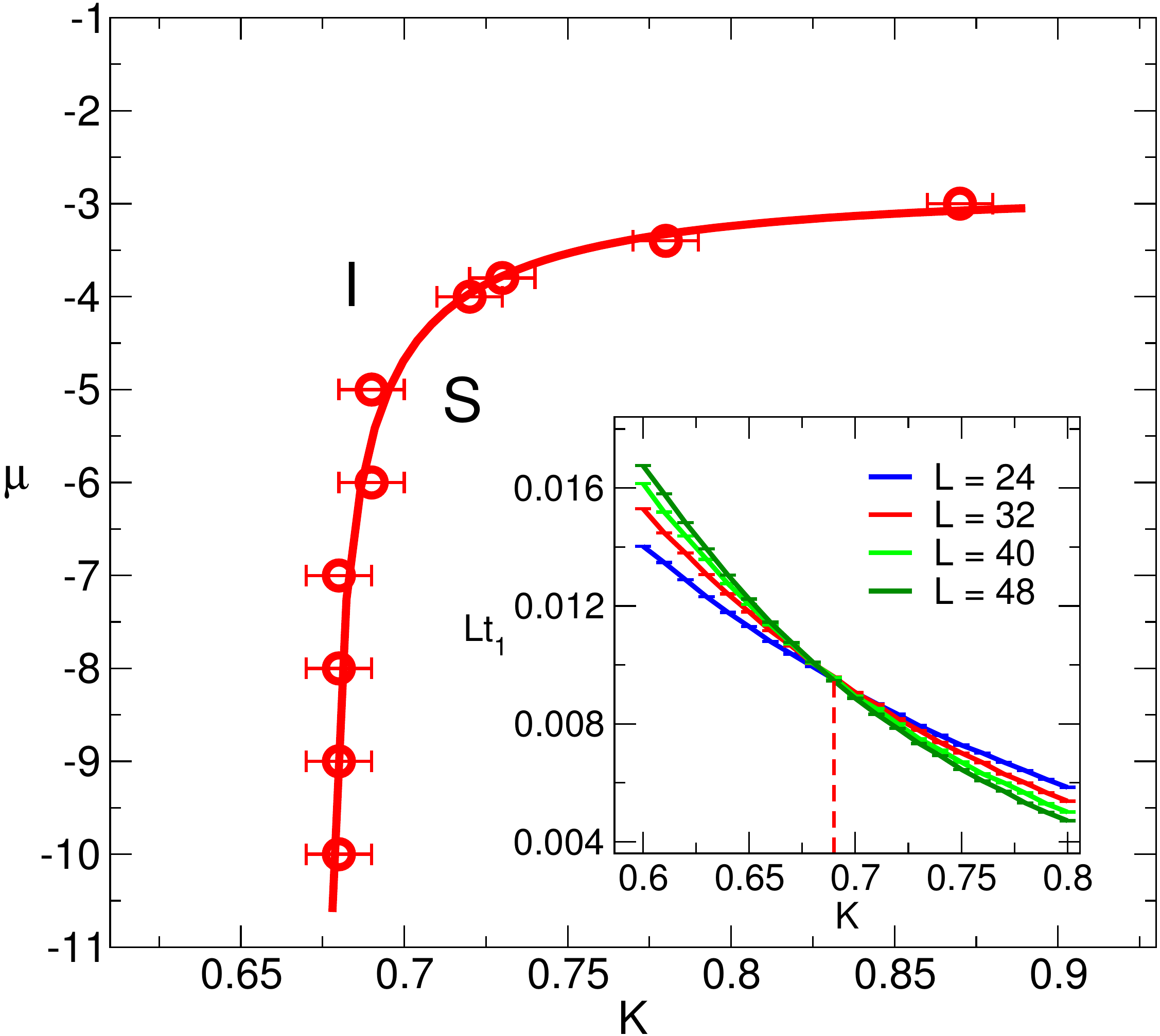}
     \caption{(Color online) A phase diagram of the model in the limit $\lambda \ll
       x_0 \ll L$ ($\lambda = 0.001$, $x_0 = 1$), in the coupling
       constant $\tilde K$ and the chemical potential $\mu$ plane.
       The $\tilde K_c$ curve is constructed from the intersection of
       $Lt_1$ vs $\tilde K$ curves for $L,\beta$ = 24, 32, 40, and
       48.
       The letter S marks the superconducting phase, and I the insulating phase.
       Inset: Example of the scaled phase slip amplitude $L t_1$ vs $K$ for 
       different $L$ intersecting at $K \approx 0.68$ for $\mu = -8$.
       \label{fig:K_vs_mu_Kc}
     }
   \end{center}
\end{figure}
Interactions are important both near the superconductor-insulator
transition and in the region of strong coherent QPS.
The signatures of a BKT transition will be most clear in the case
of a purely linear
dispersion (see \Eq{eq:epsilon}), which is formally obtained by taking
$\lambda \ll \xi$, since this gives a purely logarithmic QPS
interaction.
In \Fig{fig:K_vs_mu_Kc}, we plot the phase diagram in this limit,
as functions of $K$ and $\mu = \ln y$.
[A similar phase diagram is obtained in the case of Josephson junction
chains, but with $K=\sqrt{E_J/E_{C_0}}$, $\lambda = \sqrt{C/C_0}$, and
$\mu \sim - \sqrt{E_J/E_C}$.]
At the transition $t_1 \sim L^{-1}$, as discussed in
Sec.~\ref{sec:SIT}, and this is used to locate the transition as the
parameter values where $L t_1$ is independent of $L$, as shown in the
inset.
The phase boundary is nearly vertical for small $\mu$ (large
$\Score$), where $K$ remains largely unrenormalized.\footnote{The critical $K$
  obtained is slightly higher than the expected $2/\pi$, due to
  subleading corrections to the asymptotic logarithmic $L$ dependence
  in \Eq{eq:V00}.}
Screening of the QPS interaction then renormalizes $K$ towards higher
values when $\mu$ increases, reflected by the bending of the phase
boundary above $\mu \approx -5$.

Note that the overall magnitude of $t_1$ is strongly suppressed when
$\lambda \gg \xi$ in the regime near the SI transition, since the QPS
action $S_0 \sim K\lambda/\xi \gg 1$, when $K \gtrsim 2/\pi$.
The phase boundary observed in experiments on wires, whose length is
smaller than or comparable to $\lambda$, is therefore more likely a
crossover between low and high QPS amplitudes, at much lower $K$.

Effects of interactions are perhaps easier to observe in the regime of
coherent QPS.  As discussed in Sec.~\ref{sec:DL}, the response of the
wire to an applied charge displacement $k$ is, in the noninteracting
limit, governed by simple cosine or sine behaviors in $E_k$, $V_k$,
and $C_k^{-1}$.  Any deviations from these forms are in this sense a
sign of interactions between QPS.

\begin{figure}[]
   \begin{center}
     \includegraphics[width = 7.5cm]{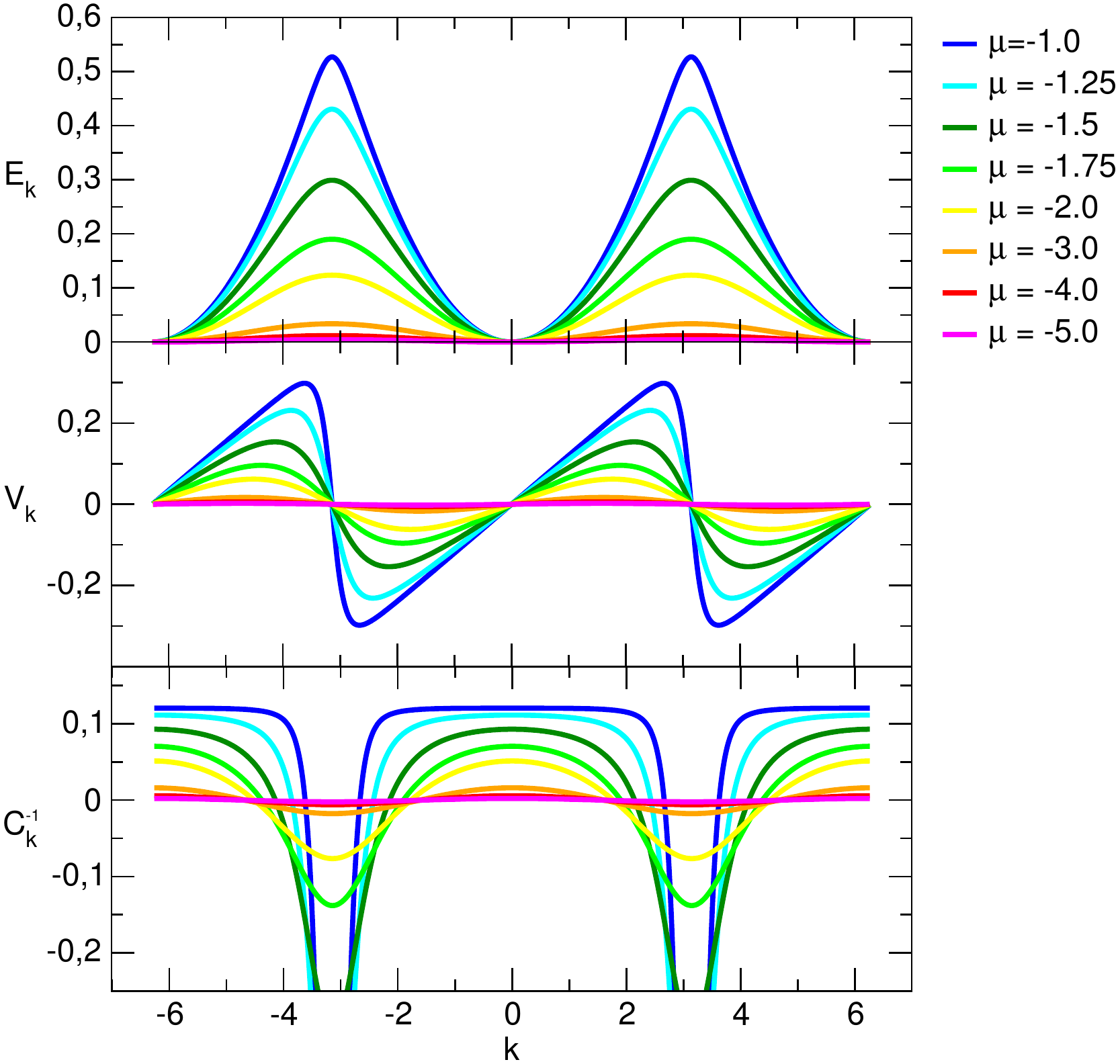}
     \caption{(Color online) $E_k$, $V_k$, and $C_k^{-1}$ for different $\mu$ at $\lambda/x_0 = 20$ and $K = 0.3$.
       System size is $10 x_0 \times 10 \tau_0$.
       \label{fig:Ek_lambda20_K0.3}
     }
   \end{center}
\end{figure}
\begin{figure}[]
   \begin{center}
     \includegraphics[width = 7.5cm]{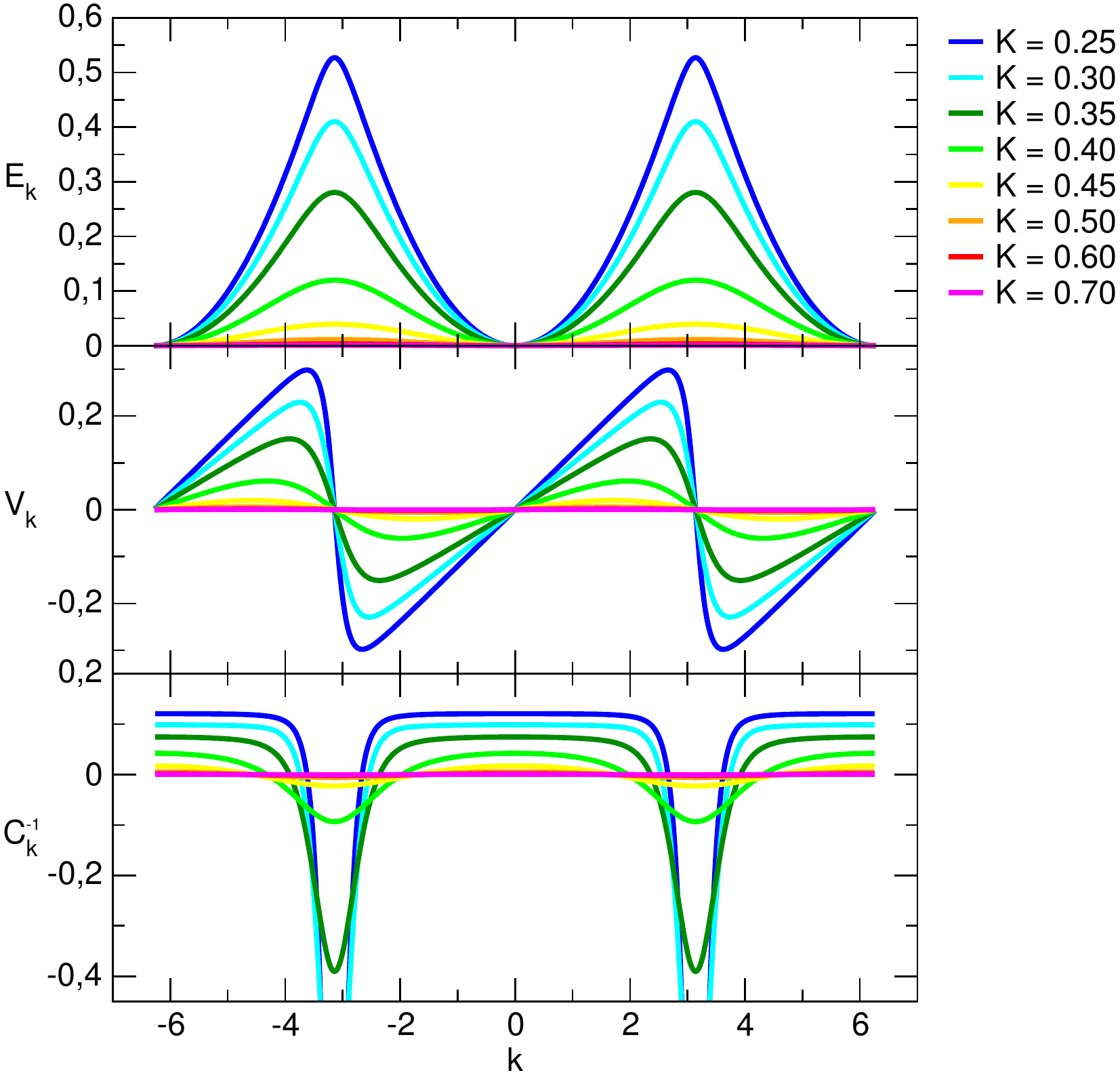}
     \caption{(Color online) $E_k$, $V_k = dE_k/dk$, and $C_k^{-1} = d^2E_k/dk^2$ for
       different $K$ at $\lambda/x_0 = 20$ and $\mu = -1$. System size
       is $10 x_0 \times 10 \tau_0$.
       \label{fig:Ek_lambda20_mu-1}
     }
   \end{center}
\end{figure}
\begin{figure}[]
   \begin{center}
     \includegraphics[width = 7.5cm]{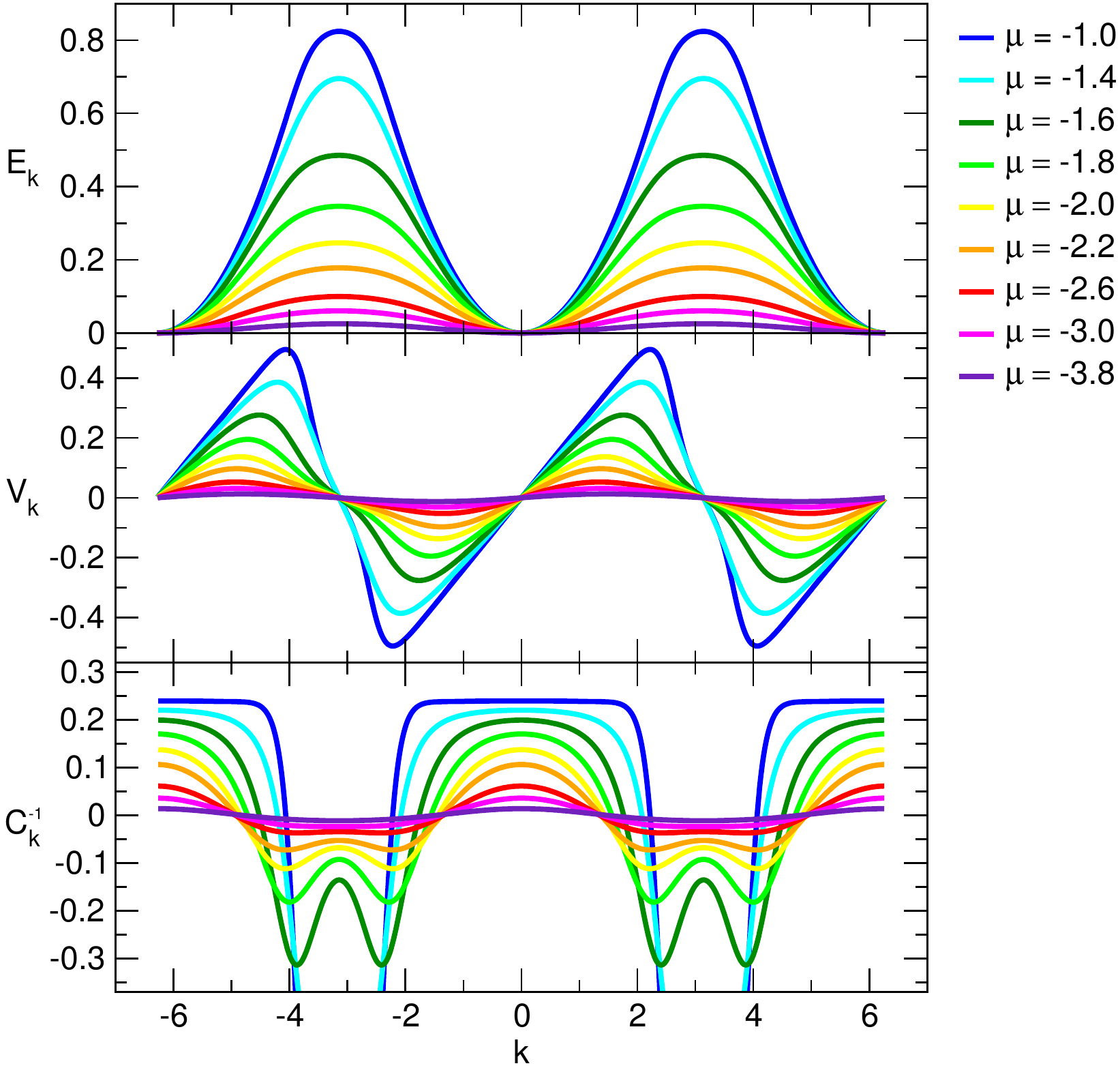}
     \caption{(Color online) $E_k$, $V_k$, and $C_k^{-1}$ for different $\mu$ at $\lambda/x_0 = 1$ and $K = 1$.
       System size is $10 x_0 \times 10 \tau_0$.
       \label{fig:Ek_lambda1.0_K1.0}
     }
   \end{center}
\end{figure}
\begin{figure}[]
   \begin{center}
     \includegraphics[width = 7.5cm]{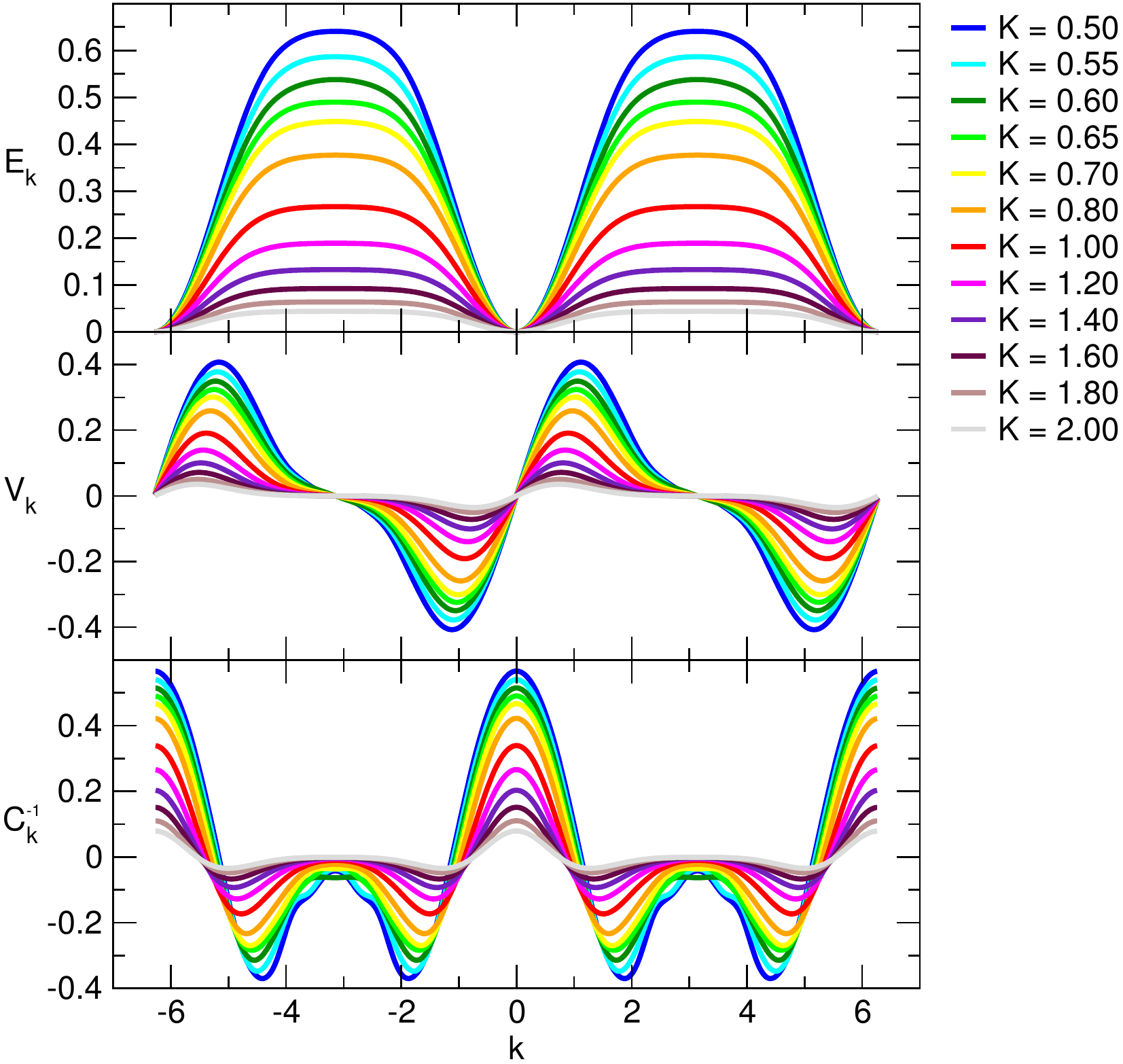}
     \caption{(Color online) $E_k$, $V_k$, and $C_k^{-1}$ for different $K$ at $\lambda/x_0 = 0.001$ and $\mu = -3$.
       System size is $10 x_0 \times 10 \tau_0$.
       \label{fig:Ek_lambda0.001_mu-3}
     }
   \end{center}
\end{figure}
\begin{figure}[]
   \begin{center}
     \includegraphics[width = 7.5cm]{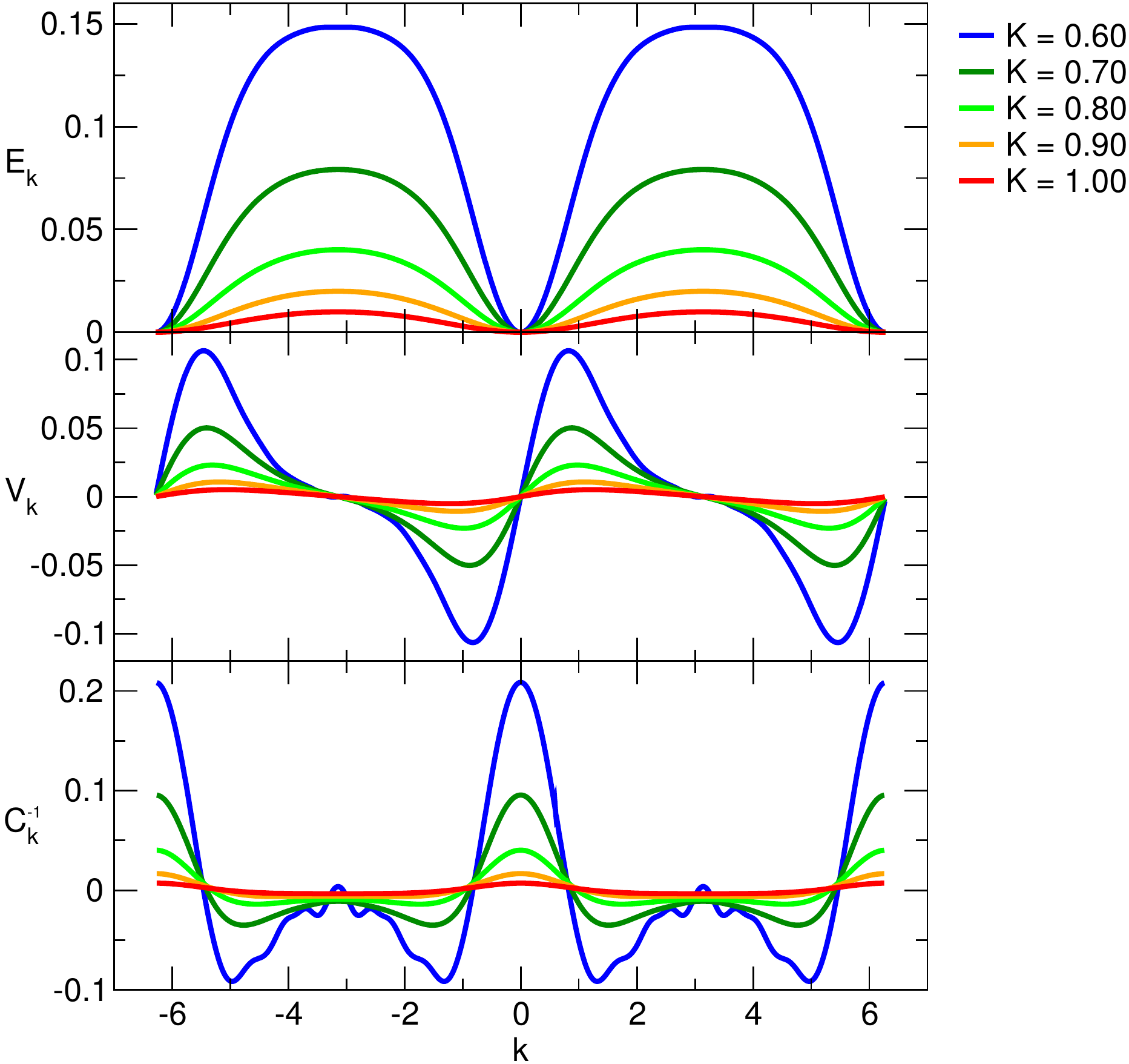}
     \caption{(Color online) $E_k$, $V_k$, and $C_k^{-1}$ for different $K$ at $\lambda/x_0 = 1$ and $\mu = -3$, for a system of size 
       $40 x_0 \times 40 \tau_0$. The small scale oscillations are due
       to statistical noise in the simulation.
       \label{fig:L40_Ek_lambda1.0_mu-3}
     }
   \end{center}
\end{figure}
In \Fig{fig:Ek_lambda20_K0.3} curves of $E_K$, $V_k$, and $C_k^{-1}$
from simulations in the limit ($\lambda/x_0 \gg 1$), show how
these quantities change from sinusoidal to
distinctly different shapes, as the chemical potential $\mu$
increases, and thus the density of QPS increases.  At the higher
densities,
$E_k$ approaches a pattern of crossing parabolas with a period
of $2\pi$ (corresponding to charge transfers of $2e$ Cooper pairs), indicative of a situation with a wildly fluctuating
superconducting phase and a well defined charge variable, conjugate to
the phase. Here, the low energy excitations of the wire are dominated
by a capacitative energy, quadratic in the charge displacement $k$. The maximum
of the sawtooth-like $V_k$ curves, tells us what voltage is required
to push a Cooper pair through the wire, i.e., the threshold voltage it
takes to overcome the Coulomb blockade in this phase. 
Further, the inverse of the effective capacitance $C_k^{-1}$
increases steadily for higher densities of QPS (larger $\mu$), as can
be seen in the bottom panel of \Fig{fig:Ek_lambda20_K0.3}, until it
reaches the limiting value $L /(C\lambda^2)$.

An almost identical evolution of the shapes of the energy, voltage, and
inverse capacitance vs $k$ can be seen in \Fig{fig:Ek_lambda20_mu-1},
where instead the coupling constant $K$ is varied and the chemical
potential $\mu = -1$ is kept constant, but with the same effect of
changing the phase slip density (small $K$ corresponds to high phase
slip density and vice versa).

Focusing on the case of a shorter charge screening length $\lambda$,
the curves can
take on more exotic shapes than discussed previously. In
\Fig{fig:Ek_lambda1.0_K1.0}, which displays simulation results for
$\lambda/x_0 = 1$ and $K = 1$ while varying $\mu$, the crossing of the
parabolas in $E_k$ from above are now distinctly more rounded and
somewhat flattened, giving the $V_k$ sawtooth curves a backward tilt.
This trend is even more pronounced in \Fig{fig:Ek_lambda0.001_mu-3},
where the charge screening length is even smaller, $\lambda / x_0 =
0.001$ and the chemical potential $\mu = -3$.  A similar behavior can
be obtained by keeping $\lambda$ fixed and going to larger system
sizes, so that $\lambda$ is still much smaller than $L$. An example of
this can be seen in \Fig{fig:L40_Ek_lambda1.0_mu-3}, where $\lambda /
x_0 = 1$ and $L = 40 x_0$.
These peculiar shapes can be understood by a simple argument: as the
charge displacement $e k /\pi$ increases from zero, a pair of positive
and negative Cooper pairs is created and dragged apart leading to an
initial steep rise of the energy.  As the separation becomes
comparable to the charge screening length $\lambda$, i.e., when
$k \gtrsim 2\pi \lambda / L$, the attractive interaction will start to
level off and eventually flatten out since there is no extra energy
required to separate the pair further.

\section{Conclusions}

We have constructed a framework for modeling and simulations of
quantum phase slips, based on a microscopic model of a superconducting
ultrathin wire.
The low-energy effective
action~\cite{GolubevZaikinPRB2001,ArutyunovGolubevZaikinPR2008} is
transformed, first, into a charge-current model and then to a gas of
interacting instantons.
Our treatment difffers from most earlier works by the inclusion of
electromagnetic fluctuations, thus enabling the tunneling of flux
quanta across the wire, between the otherwise 
degenerate flux states.

With this we are able to calculate, via Monte Carlo simulations, the
QPS tunneling amplitude beyond the dilute instanton gas approximation,
without any restriction of the QPS density, and study the effects of
instanton interactions.  The regime of validity of the noninteracting
instanton gas approximation is found to be limited to quite low
densities, see \Fig{fig:t1_and_t1D_vs_LoverRN}.

By tuning the QPS action $S_0$ and coupling constant $K$ (both
dependent on the thickness $s$ of the wire), it is possible to go from
a superconducting state with few incoherent QPS to an insulating state
with coherent QPS.
In the coherent regime, the linear response to an applied voltage is
capacitative.
Only when the voltage exceeds a certain threshold voltage will current
start to flow.  We have calculated the corresponding voltage-charge
displacement relation $V(k)$ and the effective (inverse) quantum capacitance
$C^{-1}(k)$ of the whole wire.
These change from simple sinusoidal relations at low QPS density into
sawtooth and other nontrivial shapes as the density increases and
hence interactions become more important.

Circuits operating in the coherent regime will in many respects be dual to
superconducting ones~\cite{MooijNazarovNature2006}.
For example, an applied constant current $I = (e/\pi) dk/dt$ would lead
to Bloch oscillations~\cite{Lehtinen2012a}, i.e., an oscillating voltage with frequency
$\nu = I/2e$, which holds the promise of defining an electric current
standard, which is currently lacking.
One may speculate that frequency locking of the oscillations
and the observation of dual Shapiro steps could be facilitated by
tuning the shape of the $V(k)$ curves.
Experimental efforts to realize devices based on QPS are under
development~\cite{Hongisto2012,HriscuNazarovPRB2011}.
A Cooper pair transistor device dual to the dc SQUID was recently
suggested~\cite{Hongisto2012}, which might admit experimental studies
of the voltage-charge relations calculated above.

The calculational methods developed here lays the foundations for
future quantitative studies of the influence of disorder, dissipation,
and temperature in superconducting nanowires and more complicated
structures and devices.

\acknowledgments

Discussions with T.~H.~Hansson and D.~Haviland are gratefully acknowledged.
The simulations were performed on resources provided by the Swedish
National Infrastructure for Computing (SNIC) at 
PDC Centre for High Performance Computing (PDC-HPC).

\bibliography{qps}
\end{document}